\documentclass[a4paper]{article}
\usepackage[english]{babel}
\usepackage[utf8x]{inputenc}
\usepackage{amsmath}
\usepackage{fixmath}
\usepackage{amsfonts}
\usepackage{amssymb}
\usepackage{tabularx}
\usepackage{dirtytalk}
\usepackage{graphicx}
\usepackage{subfigure}
\usepackage{hyperref}
\usepackage{bbm}
\usepackage[colorinlistoftodos]{todonotes}
\usepackage{geometry}
\usepackage{orcidlink}

\usepackage[sorting=none, url=false, doi = false]{biblatex}
\usepackage{dirtytalk}
\usepackage{color}

\title{QML-FAST - A Fast Code for low-$\ell$ Tomographic Maximum Likelihood Power Spectrum Estimation}

\usepackage{authblk}

\author[1]{Yurii Kvasiuk\, \orcidlink{0009-0002-4720-1320}} 
\author[1]{Anderson Lai\, \orcidlink{0000-0003-2741-4556}} 
\author[1,2]{Moritz M\"unchmeyer\, \orcidlink{0000-0002-3777-7791}} 
\author[3]{Kendrick M. Smith, \orcidlink{0000-0002-3777-7791}} 

\affil[1]{Department of Physics, University of Wisconsin-Madison, Madison, WI 53706, USA}
\affil[2]{NSF-Simons AI Institute for the Sky (SkAI), Chicago, IL 60611, USA}
\affil[3]{Perimeter Institute for Theoretical Physics, Waterloo, ON N2L 2Y5, Canada}

\date{\today}

\begin{document}

\maketitle

\begin{abstract}
We present a novel implementation for the quadratic maximum likelihood (QML) power spectrum estimator for multiple correlated scalar fields on the sphere. Our estimator supports arbitrary binning in redshift and multipoles $\ell$ and includes cross-correlations of redshift bins. It implements a fully optimal analysis with a pixel-wise covariance model. We implement a number of optimizations which make the estimator and associated covariance matrix computationally tractable for a low-$\ell$ analysis, suitable for example for kSZ velocity reconstruction or primordial non-Gaussianity from scale-dependent bias analyses. We validate our estimator extensively on simulations and compare its features and precision with the common pseudo-$C_\ell$ method, showing significant gains at large scales. We make our code publicly available\footnote{https://github.com/ykvasiuk/qmlfast}. In a companion paper, we apply the estimator to kSZ velocity reconstruction using data from ACT and DESI Legacy Survey and construct full set of QML estimators on 40 correlated fields up to $N_{\text{side}}= 32$ in timescale of an hour on a single 24-core CPU requiring $<256\ \mathrm{Gb}$ RAM, demonstrating the performance of the code.
\end{abstract}

\tableofcontents

\section{Introduction}
A central statistic of cosmological theory and data analysis are power spectra and cross-power spectra of various fields. For this reason, there is a large toolkit of different power spectrum estimators, including the popular and fast (but not optimal) pseudo-$C_\ell$ method \cite{Wandelt:2000av}. Optimal power spectrum estimation, with statistically minimal error bars, remains challenging in practice. The Quadratic Maximum Likelihood (QML) estimator was introduced in \cite{Tegmark_1997} as a method to optimally estimate the power spectrum of the CMB temperature anisotropy and extended to include polarization power spectra in \cite{Tegmark_2001}. It was shown that this estimator is an optimal and unbiased estimator of the power spectrum multipoles given the pixel map with an arbitrary mask. The estimator was applied to study large-scale WMAP polarization spectra in \cite{Gjerl_w_2015}. An extension of the formalism, xQML, that allows correlating different CMB datasets was presented in \cite{Vanneste:2018azc}. A more optimized implementation of the standard QML estimator was introduced in \cite{Bilbao_Ahedo_2021}. There, the authors discuss how the sparsity of the basis matrices of the signal covariance decomposition and precomputation and reuse of certain elements can speed up the Fisher matrix calculation, which is the most computationally intensive part. In \cite{Pen_2003}, an iterative procedure is presented to avoid the expensive operation of directly inverting the covariance matrix. The applications of QML to weak lensing surveys were discussed in \cite{Maraio_2023}, where the authors introduced several approaches to deal with $\ell^6$ scaling of the method, such as evaluating an approximate Fisher matrix with finite differences. In a more recent galaxy weak lensing study~\cite{Terawaki:2025:weak_lensing}, the computational cost of quadratic estimators is reduced by assuming a diagonal fiducial covariance and sampling the Fisher matrix through Monte-Carlo simulation. In addition to weak lensing analysis,~\cite{Alonso:2024knf} applied a QML estimator to look for the production rate of gravitational waves through the galaxy-gravitational wave cross power spectrum. A hybrid, near-optimal estimator, that combines QML for low multipoles and pCl for high multipoles was developed in \cite{Efstathiou_2004, Efstathiou:2006eb}. All the previous implementations of the field-level QML estimator were developed for the analyses where one only has up to 3 correlated fields. In the typical CMB analysis one has $T,E,$ and $B$ fields, or in \cite{Alonso:2024knf} authors consider correlations of galaxy and gravitational wave background maps. The need for a multi-field version of the QML is motivated by the fact that it's an ideal tool for the low-$\ell$ power spectrum analysis and can be used in the context of the photometric galaxy surveys, where galaxy samples are usually binned in redshift bins with a width proportional to the photometric redshift error. In this work, we generalize the QML framework to an arbitrary number of correlated fields. We introduce further optimization techniques, such as symmetries of the building blocks of the Fisher matrix and the basis of real spherical harmonics, in addition to the ones already discussed in \cite{Bilbao_Ahedo_2021} and provide both efficient and user-friendly $\texttt{python}$ implementation. Our resulting code is several orders of magnitude faster than a naive implementation and will be useful to analyze large-scale data from upcoming photometric galaxy surveys and CMB experiments. This paper presents the features of our QML estimator and validates it with simulations, while the companion paper~\cite{Lai2025kszpaper} applies it to the case of kSZ velocity reconstruction with ACT and Desi Legacy Survey data.

This paper is organized as follows. We give an overview of the QML formalism, generalized to multiple fields in the Section \ref{sec:qml}, where we also discuss the conditions of unbiasedness, optimality, relations to the field-level likelihoods, removal of the auto-powers from cross-powers, and removal of unwanted $\ell$-modes. In the Section \ref{sec:optimization}, we discuss the details of efficient computational implementation and parametric complexity. The Section \ref{sec:validation} provides the details of numerical validation and performance comparison to the pseudo-$C_{\ell}$ method. We conclude in the Section \ref{sec:conclusion}.

\section{Review of the QML Formalism for Power Spectrum Estimation}
\label{sec:qml}
We provide a brief review of the quadratic maximum likelihood power spectrum estimator \cite{Tegmark_1997}, generalized to an arbitrary number of correlated scalar Gaussian random fields. Let ${\pmb\phi} = \{\phi_\alpha\}$ represent the pixel-space fields, the power spectra of which we want to estimate. Assuming $N_f$ fields and $N_p$ is a number of unmasked pixels, the total pixel-space covariance is $N_f\times N_f$ matrix of $N_p\times N_p$ blocks. We decompose the signal term in the power spectrum basis:
\begin{equation}
\label{eq:cov_decomposition}
    \langle\pmb{\phi\phi^T}\rangle = \mathbb{S+N} = \sum_{\{A\}}P_Ac_A +\mathbb{N}
\end{equation}
Motivated by cosmological applications, we assume each field to be a 2-d scalar field on a sphere, however, the formalism is easily generalizable to arbitrary fields. Then, $c_A$ is a $\ell$-th multipole of the cross power spectrum of the pair $\alpha\beta$ and
\begin{equation}
    P_A \equiv \frac{\partial \mathbb{C}}{\partial c_A}
\end{equation}
is the basis element. If $A=(\alpha,\beta,\ell)$, then block matrix $P_A$ is $P_A= \delta_{\alpha\beta}\otimes P^\ell_{ij}$ where $P^{\ell}_{ij} = \frac{2\ell+1}{4\pi}\mathcal{P}^\ell(\hat{\mathbf{n}}_i \cdot \hat{\mathbf{n}}_j)$, and $\mathcal{P}^{\ell}(\hat{\mathbf{n}}_i\cdot \hat{\mathbf{n}}_j)$ is a Legendre polynomial of the order $\ell$ of the cosine of the angle between unit-vectors pointing in the direction of $i$-th and $j$-th pixels. Then the estimator of $c_A$ is given by
\begin{equation}
\label{eq:qml}
    \hat{c}_A = (F^{-1})_{AB}\left[\pmb{\phi}^TQ_B\pmb{\phi}-n_B\right]
\end{equation}
Where 
\begin{equation}
    Q_A = \frac{1}{2}C^{-1}P_AC^{-1},\quad n_A = \texttt{Tr}[NQ_A] 
\end{equation}
And 
\begin{equation}
\label{eq:fisher_def}
    F_{AB} = \frac{1}{2}\texttt{Tr}[\mathbb{C}^{-1}\mathbb{C}_{,c_A}\mathbb{C}^{-1}\mathbb{C}_{,c_B}]\equiv \frac{1}{2}\texttt{Tr}[\mathbb{C}^{-1}P_A\mathbb{C}^{-1}P_B]
\end{equation}
This estimator is unbiased and optimal according to the Cramer-Rao lower bound:
\begin{equation}
    \langle\hat{c}_A\rangle = c_A, \quad \langle(c_A-\hat{c}_A)( c_B-\hat{c}_B)\rangle = (F^{-1})_{AB}
\end{equation}

\subsection{Mismodelling the data covariance}
The last statement assumes that our fiducial model covariance used for the Fisher matrix calculation equals the data covariance. This is not always the case, as often we are interested in the parameter inference where the fiducial covariance is only assumed to be close to the true data covariance. However, the estimator remains unbiased, even if we use the wrong signal covariance (but not the noise covariance!). Let's denote the mismodeled quantities with tildes, eg. $\tilde{\mathbb{C}} = \tilde{\mathbb{S}} + \tilde{\mathbb{N}}$, and $\tilde{Q}_A = \frac{1}{2}\tilde{\mathbb{C}}^{-1}P_A\tilde{\mathbb{C}}^{-1}$. Let's define \begin{equation}
\label{eq:y}
\hat{y}_A \equiv F_{AB}\hat{c}_B = \texttt{Tr}[Q_A(\pmb{\phi}\pmb{\phi^T}-\mathbb{N})]
\end{equation}
Then it follows that
\begin{equation}
    \langle\tilde{y}_A\rangle = \texttt{Tr}[\tilde{Q}_A(\mathbb{S}+(\mathbb{N}-\tilde{\mathbb{N}}))] = \sum_{AB}c_B\texttt{Tr}[\tilde{Q}_AP_B] + \texttt{Tr}[\tilde{Q}_A(\mathbb{N}-\tilde{\mathbb{N}})]
\end{equation}
But $\texttt{Tr}[\tilde{Q}_AP_B] = \tilde{F}_{AB}$. So that the estimator $\tilde{\hat{c}}$
\begin{equation}
    \langle (\tilde{F}^{-1})_{AB}\tilde{y}_B\rangle \equiv \langle \tilde{\hat{c}}_A\rangle = c_A + (\tilde{F}^{-1})_{AB}\texttt{Tr}[\tilde{Q}_B(\mathbb{N}-\tilde{\mathbb{N}})]
\end{equation}
is unbiased unless the noise is mismodeled. However, the resulting estimator is no longer optimal, as we show in  Appendix \ref{app:mismod_var}.

\subsection{Connection to the field-level likelihood}
Here, we want to draw a parallel between using the QML estimator and maximizing the field-level likelihood. Consider a typical cosmological parameter inference problem. We are interested in a set of parameters $\Omega$, having measured data $\pmb{\phi}$ that is assumed to be a mean-zero Gaussian random field with a parameter-dependent covariance. Then MLE $\Omega^*$ is typically found by solving
\begin{align}
\Omega^* &= \mathrm{argmax}_{\Omega}
  \ln \mathcal{L}_{field}(\pmb\phi|\Omega) = \\
  &=\mathrm{argmin}_{\Omega} \left[\frac{1}{2}\pmb{\phi^T}(\mathbb{S}(\Omega)+\mathbb{N})^{-1}\pmb\phi +\frac{1}{2}\ln\det(\mathbb{S}(\Omega)+\mathbb{N})\right]
\end{align}
Define the score function $s(\Omega)$ by $s({\Omega})\equiv\partial_\Omega\ln{\mathcal{L}}$. Let's expand it at some $\Omega_0$ up to the first order. At the optimum we have $s({\Omega^*}) = 0$:
\begin{equation}
    0 = s({\Omega_0}) + (\hat{\Omega}^*-\Omega_0)\partial_{\Omega}s|_{\Omega_0}
\end{equation}
If we substitute $\partial_{\Omega}s\rightarrow\langle\partial_{\Omega}s|_{\Omega_0}\rangle=\langle\partial^2_{\Omega}\mathcal{L}|_{\Omega_0}\rangle = -F_0^{\Omega\Omega}$, 
we get
\begin{equation}
\label{eq:nr_update}
    \hat{\Omega}^* = \Omega_0 + (F_0^{\Omega\Omega})^{-1}s(\Omega_0)
\end{equation}
It is exactly a Newton-Raphson update step, and the QML estimator is equivalent to the first update step. For the Gaussian score, we have
\begin{equation}
    \partial_{\Omega}\ln\mathcal{L} = \frac{1}{2}\texttt{Tr}[\mathbb{C}^{-1}(\partial_{\Omega}\mathbb{C})\mathbb{C}^{-1}\pmb{\phi\phi}^T]-\frac{1}{2}\texttt{Tr}[\mathbb{C}^{-1}(\partial_{\Omega}\mathbb{C})]
\end{equation}
Identifying $\Omega$ with $c_A$, we have then 
\begin{equation}
\texttt{Tr}[\mathbb{C}^{-1}P_A] = \texttt{Tr}[\mathbb{C}Q_A] = \texttt{Tr}[Q_A(\mathbb{S+N})] = \texttt{Tr}[Q_A\mathbb{N}] + F_{AB}c_B,
\end{equation}
so that the equation \eqref{eq:nr_update} becomes identical to the equation \eqref{eq:qml}.
Another useful property is that in the approximation where QML estimates are Gaussian, which is true to leading order in $1/l$ by the central limit theorem, their covariance is $\mathrm{cov}^{-1}(\hat{c})_{AB} = F_{AB}$, so that the information content of the power-spectrum likelihood 
\begin{equation}
    \ln\mathcal{L}_{ps}(\hat{c}_A(\pmb{\phi})|\Omega) \cong -\frac{1}{2}(\hat{c}_A-c_A(\Omega))\mathrm{cov}(\hat{c})^{-1}_{AB}(\hat{c}_{B}-c_B(\Omega)) 
\end{equation}
is equal to the field-level one, as
\begin{equation}
    F^{ps}_{\Omega\Omega} =\langle\partial_\Omega\ln\mathcal{L}^{ps}\partial_\Omega\ln\mathcal{L}^{ps}\rangle =\frac{\partial c_A}{\partial \Omega}F_{AB}\frac{\partial c_B}{\partial\Omega} = \frac{1}{2}\mathrm{Tr}[\mathbb{C}^{-1}(\partial_\Omega\mathbb{C})\mathbb{C}^{-1}(\partial_\Omega\mathbb{C})] = F^{field}_{\Omega\Omega}
\end{equation}
This property can be useful for the exact Fisher forecasting.

\subsection{Binned estimator}\label{subsec:Bandpowered estimator}
To get the estimator, we need to calculate 
\begin{equation}
    \hat{c}_{A} = (F^{-1})_{AB}\hat{y}_{B}
\end{equation}
In the case when the sky coverage is sufficiently small, $(F_{AB})^{-1}$ might not exist. The minimum resolution width in $\ell$-space is inversely proportional to the angular separation \cite{Tegmark_1996}:
\begin{equation}
    \Delta\ell \sim \frac{1}{\Delta\theta}
\end{equation}
where $\Delta\theta$ is the smallest angular separation of the sky patch in radians. For simplicity, we consider a one-field case and introduce a linear binning operator $\mathcal{P}$:
\begin{equation}\label{eq: bandpowering y_ell}
    \hat{y}_b = \mathcal{P}_{b\ell}\hat{y}_{\ell}
\end{equation}
Then we can define "unbinning" operator $\mathcal{S}$ to be the right pseudo-inverse $\mathcal{S} = \mathcal{P}^T(\mathcal{P}\mathcal{P}^T)^{-1}$ such that
\begin{equation}\label{eq: bandpowering pinv}
    \hat{y}_{b'} = \mathcal{P}_{b'\ell}\mathcal{S}_{\ell b} \hat{y}_{b}
\end{equation}
and assume that
\begin{equation}
    c_{\ell} \approx \mathcal{S}_{\ell b}c_{b}
\end{equation}
Then
\begin{equation}
    \langle y_{b}\rangle = \mathcal{P}_{b\ell}F_{\ell\ell'}\mathcal{S}_{\ell'b'}c_{b'}
\end{equation}
For the binned estimator we have
\begin{equation}\label{eq: bandpowering QML estimate}
    \hat{c}_{b} = (\mathcal{P}F\mathcal{S})^{-1}_{bb'}\mathcal{P}_{b'\ell}\hat{y}_{\ell}
\end{equation}

and for the variance
\begin{equation}\label{eq:Bandpowered Fisher}
    \langle(c_b-\hat{c}_b)( c_{b'}-\hat{c}_{b'})\rangle = (S^TFS)^{-1}_{bb'} \equiv F^{-1}_{bb'}.
\end{equation}

\subsection{Deprojection for cross-power spectra}
\label{subsec:deprojx}
QML estimator is minimum-variance by construction. However, for the case of multiple fields, this results in the presence of the autopowers in the estimator of the cross power
(\cite{Tegmark_2001}, Sec. III D). This is often undesired, for example when auto-powers are affected by large calibration systematics. In the case when we are interested in measuring the cross-correlation, we naturally would like to ensure that the estimator of cross-powers is constructed only from the cross-products or the corresponding fields $\hat{C}_{XY} \supset XY$. This is achieved by setting the corresponding entries $C_{XY}$ in the fiducial covariance to zero. We note here that this ad-hoc prescription is natural when one is concerned with the cross-power detection SNR. For example, when we have two fields (or two groups of fields) $\phi_X$ and $\phi_Y$, we introduce an amplitude $\mathcal{A}$ in the total covariance:
\begin{equation}
    \mathbb{C}=\begin{pmatrix}
        C_{XX} & \mathcal{A}C_{XY} \\
        \mathcal{A}C_{YX} & C_{YY}
    \end{pmatrix}
\end{equation}
Then for the rejection of the null hypothesis that the fields are uncorrelated $H_0:\{A=0\}$, we usually need to evaluate $\sigma^2_\mathcal{A}\bigg|_{\mathcal{A}=0} = F^{-1}_{\mathcal{A}\mathcal{A}}\bigg|_{\mathcal{A}=0}$ setting the cross-powers equal to zero in the fiducial covariance explicitly. We also note that setting the off-diagonal blocks of the total covariance to zero can be leveraged to further reduce the memory consumption and the time to calculate the Fisher matrix. We elaborate on this in \ref{subsec:fisher_eval}.

\subsection{Deprojection of the unwanted multipoles}\label{subsec:deproj_ell}
In usual applications, we often need to restrict our analysis to the $\ell$-modes only for some $\ell>\ell_{min}$ (for example the monopole and dipole are usually excluded from CMB analysis). It was argued in \cite{Tegmark_1997}, that in the 1 field case, one can define the "pseudo-inverse" matrix $M=\pi[\pi C\pi^T+\eta ZZ^T]^{-1}\pi^T$, where $Z$ is constructed from the orthonormalized columns corresponding to the unwanted multipoles, $Z^TZ = 1$; and $\pi = 1-ZZ^T$ - a projector to the orthogonal subspace. The matrix $M$ is used everywhere instead of $C^{-1}$ and the result is independent of $\eta$. The same argument extends to the case of general $N$ fields. We define \begin{equation}
    \mathrm{\Pi} = \mathbbm{1}_{N_f} \otimes \pi
\end{equation}
Here, the $\mathbbm{1}_{N_f}$ is an identity matrix of the size $N_f \times N_f$. Then the matrix $\mathbb{M}$ is defined as follows:
\begin{equation}
    \mathbb{M} = \mathrm{\Pi}\left[\mathrm{\Pi}\mathbb{C}\mathrm{\Pi}+\eta(\mathbbm{1}_{N_b}\otimes ZZ^T)\right]^{-1}\mathrm{\Pi}
\end{equation}
We use the matrix $\mathbb{M}$ instead of $\mathbb{C}^{-1}$ in the definition of the Fisher matrix and the estimator. This procedure still results in the optimal estimator for the remaining multipoles. In the Appendix \ref{app:pi}, we provide the justification for this statement.

\subsection{An alternative approach}
There's another approach to mode projection that is simpler than the pseudo-inverse. Conceptually, we project out modes by assigning them large variance, i.e.\ replacing $C^{-1}$ by the matrix:
\begin{equation}
C^{-1} \rightarrow \big[ C + \alpha ZZ^T \big]^{-1}
  \hspace{1cm} \mbox{where } \alpha \gg 0
\end{equation}
Here, $Z$ is a matrix with one column per mode that we want to project. We can take the limit $\alpha \rightarrow \infty$ as follows:
\begin{align}
\lim_{\alpha \rightarrow \infty} 
    \big[ C + \alpha ZZ^T \big]^{-1}
&= \lim_{\alpha \rightarrow \infty} 
    \big[ C^{-1} - C^{-1} Z \big( Z^T C^{-1} Z + \alpha^{-1} 1 \big)^{-1} Z^T C^{-1} \big] \nonumber \\
&= C^{-1} - C^{-1} Z \big( Z^T C^{-1} Z \big)^{-1} Z^T C^{-1}
\end{align}
where we used the Sherman-Woodburry formula in the first line.
Therefore, to project out modes, we make the replacement:
\begin{equation}
C^{-1} \rightarrow C^{-1} - C^{-1} Z \big( Z^T C^{-1} Z \big)^{-1} Z^T C^{-1}
\end{equation}
where the $\alpha$ parameter no longer appears.
This formalism has the following advantages. It is conceptually simpler and manifestly equivalent to ``mode scaling'' (this was tested empirically in \S\ref{sec:validation}). It is simpler to implement since it does not require orthogonalizing $Z$ (i.e.\ $Z^TZ=1$ is not assumed), constructing the projection operator $\Pi$, etc. It's clear that this formalism carries over to the single-field and multi-field case. There are no assumptions on the matrix $Z$, so in the multifield case we take $Z$ to be a matrix with $(l_{\rm min}^2 N_f)$ columns.
\section{Computational Optimization}
\label{sec:optimization}

\subsection{Evaluating the Fisher Matrix}
\label{subsec:fisher_eval}

The most computationally intensive part of the QML estimation procedure is the calculation of the Fisher matrix, given by the equation \eqref{eq:fisher_def}. Since matrix multiplication scales approximately as $\mathcal{O}(N^3)$ with the number of elements, the total naive computational complexity makes it practically almost impossible to compute for realistic tasks. For the given $N_f,N_p$ and $\ell_{max}$ the computational complexity is $\mathcal{O}(N^4_f\ell_{max}^2)\times\mathcal{O}(N^3_fN^3_p)$. However, the covariance decomposition basis elements $P_A$ are sparse and have different symmetries that one can use to make the task computationally feasible. Let's first make the trace over field indices explicit:
\begin{equation}
    \texttt{Tr}[C^{-1}P_AC^{-1}P_B] = \sum^{N_f}_{\alpha\beta\gamma\delta}\texttt{tr}[(C^{-1})_{\alpha\beta}P^A_{\beta\gamma}(C^{-1})_{\gamma\delta}P^B_{\delta\alpha}]
\end{equation}
Here, $\texttt{tr}$ is used to denote trace over the pixels. One can notice that $P$ can be represented as a Kronecker product: 
\begin{equation}
P^A_{\gamma\delta} = P^{(\alpha,\beta,\ell)}_{\gamma\delta} = K^{(\alpha\beta)}_{\gamma\delta}\otimes P_{\ell}
\end{equation}
Here, $(P_\ell)_{ij} = \frac{2\ell+1}{4\pi}\mathcal{P}_{\ell}(\hat{\bf{n}}_i\cdot\hat{\bf{n}}_j)$, where $P_\ell(\hat{\bf{n}}_i\cdot\hat{\bf{n}}_j)$ is a Legendre polynomial of a degree $\ell$ of the cosine of the angle between the unit vectors pointing towards $i$-th and $j$-th pixels and round brackets $(\alpha\beta)$ represent ordering between the fields. We introduced the object $K^{(\alpha\beta)}_{\gamma\delta}$ that represents the sparsity of the basis elements. It's equal to $1$ only if $\gamma=\alpha,\ \delta=\beta$ or $\gamma=\beta,\ \delta=\alpha$ and $0$ otherwise. Then we can rewrite the expression 
\begin{equation}
F_{AB} = F_{(\rho\epsilon\ell_1)(\kappa\sigma\ell_2)} = \frac{1}{2}K^{(\rho\epsilon)}_{\beta\gamma}K^{(\kappa\sigma)}_{\delta\alpha}\texttt{tr}[(C^{-1})_{\alpha\beta}P_{\ell_1}(C^{-1})_{\gamma\delta}P_{\ell_2}] \equiv \frac{1}{2}K^{(\rho\epsilon)}_{\beta\gamma}K^{(\kappa\sigma)}_{\delta\alpha}\texttt{t}^{\ell_1\ell_2}_{\alpha\beta\gamma\delta}
\end{equation}
We can further decompose the trace using the addition theorem for the Legendre polynomials $\mathcal{P}_{\ell}(\hat{\bf{n}}_i\cdot\hat{\bf{n}}_j) = \frac{2\ell+1}{4\pi}\sum^{\ell}_{m=-\ell}Y^{*}_{\ell m}(\hat{\bf{n}}_i)Y_{\ell m}(\hat{\bf{n}}_j)$. For practical purposes, however, it's useful to define this decomposition in terms of the real basis. This is important because floating-point matrix operations with complex numbers take $\sim4$ times longer on modern architectures then with real numbers and require $\times 2$ the amount of RAM.

\[
Y_{\ell m}^{\mathrm{real}}(\theta,\phi)=
\begin{cases}
\dfrac{(-1)^m Y_{\ell m} + Y_{\ell,-m}}{\sqrt{2}}, & m>0, \\[0.6em]
Y_{\ell 0}, & m=0, \\[0.6em]
\dfrac{(-1)^m Y_{\ell,|m|} - Y_{\ell,-|m|}}{i\sqrt{2}}, & m<0~.
\end{cases}
\]
Then we can further simplify 
\begin{equation}
\label{eq:t_tensor}
    \texttt{t}^{\ell_1\ell_2}_{\alpha\beta\gamma\delta} = \sum_{m_1m_2} (C^{-1})_{\alpha\beta}^{\ \ ij}Y^{\ell_1m_1}_{j}Y^{\ell_1m_1}_{k}(C^{-1})_{\gamma\delta}^{\ \ kl}Y^{\ell_2m_2}_{k}Y^{\ell_2m_2}_{i} = \texttt{tr}_m \left[{}^{\ell_2\ell_1}C^{-1}_{\alpha\beta}{}^{\ell_1\ell_2}C^{-1}_{\gamma\delta}\right]
\end{equation}
In the expression above, there's no summation over $\ell_1,\ell_2$. We defined 
${}^{\ell_2\ell_1}(C^{-1})^{m_2m_1}_{\alpha\beta} = Y^{\ell_2m_2}_i(C^{-1})^{ij}_{\alpha\beta}Y^{\ell_1m_1}_j$ and the sum over pixel indices $i,j$ is implied. The benefit of this decomposition is tremendous. Typically, $N_p \propto \ell^2_{max}$ and we reduced trace over pixels to trace over $m$. From $\mathcal{O}(N_p^2)$ operations (if we precomputed the products of $C^{-1}_{\alpha\beta}P_{\ell}$) we now need to do $\mathcal{O}(\ell_{max}^2)$ operations (given that we precomputed the products $YC^{-1}Y$ before). We can further reduce the number of computations and memory by exploiting the symmetries of $\texttt{t}^{\ell_1\ell_2}_{\alpha\beta\gamma\delta}$. First we note that we can flip the first pair of field indices with the second one and simultaneously flip multipole indices. The second symmetry property follows from the fact that the covariance is a symmetric matrix: $(C^{-1})^{\alpha\beta}_{ij} = (C^{-1})^{\beta\alpha}_{ji}$. We can then flip simultaneously the multipole indices and indices in the first and second pair with each other. To conclude:
\begin{equation}
\label{eq:t_tensors}
\texttt{t}^{\ell_1\ell_2}_{\alpha\beta\gamma\delta} = \texttt{t}^{\ell_2\ell_1}_{\gamma\delta\alpha\beta} = \texttt{t}^{\ell_1\ell_2}_{\delta\gamma\beta\alpha} 
\end{equation}
 The use of the mentioned symmetries is practically important, because one needs to compute and store only the upper- (or lower-) triangular part of $YC^{-1}Y$ contractions. Some of the mentioned techniques were already discussed in \cite{Bilbao_Ahedo_2021, Gjerl_w_2015} for the case of CMB temperature and polarization power spectrum. However, here we generalize it to the case of an arbitrary number of fields and further improve by using explicit symmetries and a real representation of spherical harmonics.

\subsection{Partial case: block-diagonal covariance}
Further simplifications are possible in the case when the total data covariance takes a block-diagonal form. This is the case when we are interested in estimates of the cross powers that don't contain auto powers, as discussed in \ref{subsec:deprojx}. First, it is faster to calculate the $\mathbb{C}^{-1}$, as we can invert the blocks independently. Secondly, certain elements of the Fisher matrix are zero, while the other ones contain fewer terms. Let's, for simplicity, consider the setup when the total covariance (and its inverse) has two diagonal blocks:
\begin{equation}
    \mathbb{C}_0 = \begin{pmatrix}
        C_{\alpha\beta} && 0 \\
        0 && C_{\dot{\alpha}\dot{\beta}}
    \end{pmatrix}
\end{equation}
where we used dotted indices to denote separate blocks. This implies $(C^{-1})_{\alpha\beta}\neq0$, $(C^{-1})_{\dot{\alpha}\dot{\beta}}\neq0$, but $(C^{-1})_{\alpha\dot\beta}=0$, so that the elements $\texttt{t}^{\ell_1\ell_2}_{\alpha\beta\gamma\delta}$, introduced in eq. \eqref{eq:t_tensor}, are zero if the dotted index is between two undotted ones and vice versa. For example, $\texttt{t}^{\ell_1\ell_2}_{\alpha\dot\beta\gamma\delta} = 0$ and $\texttt{t}^{\ell_1\ell_2}_{\alpha\dot\beta\gamma\dot\delta} = 0$, but $\texttt{t}^{\ell_1\ell_2}_{\dot\alpha\dot\beta\gamma\delta} \neq 0$.

\subsection{Optimized implementation of $\ell$ binning}
To compute the binned estimates as in Sec \ref{subsec:Bandpowered estimator}, one still needs to calculate the full Fisher matrix first. It's inefficient from a computational performance perspective, as it's the most time-consuming part. We can gain an additional improvement by starting from the binned basis. For simplicity, we discuss the one-field case, since this simplifications concerns the $\ell$-part of the calculation and trivially extends to the multifield case. The derivation of the QML estimator starts from the decomposition of the pixel-space covariance in the set of basis elements, as specified in the equation \eqref{eq:cov_decomposition}. However, one can instead decompose in the binned basis:
\begin{equation}
    \mathbb{S} = \sum_{\ell}c_\ell P_{\ell} \approx \sum_b c_bP_b = \sum_{b,\ell}c_\ell\mathcal{P}_{\ell b}P_b 
\end{equation}
such that the new basis is related to the old one as follows 
\begin{equation}
    P_b = \mathcal{S}^TP_{\ell}
\end{equation}
where the operator $\mathcal{S}$ was defined in the section \ref{subsec:Bandpowered estimator}. Then, to make it computationally tractable, one needs to find an analog of the addition theorem for the new basis elements $P_b \sim \sum_{m_b}Y_{bm_b}Y_{bm_b}$. It is readily found as follows. Since for each $P_\ell$ we have $\text{rank}\ P_\ell = 2\ell+1$, by subadditivity of ranks, $\text{rank}\ P_b \leq \sum_{\ell \in \{b\}} (2\ell+1)$. Therefore, we can compute a new set of vectors by directly solving for the eigenvalues of the matrix $P_b$. We note that usually $\max \text{rank}\ P_b \ll N_\text{p}$ and for practical purposes, we can use randomized PCA solvers, specifying the corresponding number of components to be the maximum rank. Since the error of randomized solvers goes as $\mathcal{O}(\lambda_{k+1})$ where $k$ is the number of the first $k$ largest eigenvalues, we don't lose any precision, while gaining in speed ($\mathcal{O}(N^2_p)$ vs. $\mathcal{O}(N_p^3)$). The total rank of the resulting $P_b$ matrices is dependent on the exact configuration of the binning scheme and the mask. In the case when $\text{rank}\ P_b = \sum_{\ell \in \{b\}} (2\ell+1)$, there's no performance gain coming from binning, since the total number of basis elements remains the same. However, when the spectrum of the eigenvalues falls sharply, we can truncate it at some $\lambda_n,\ n<k$.

\subsection{Parametric complexity and implementation details}
Our implementation is in $\texttt{python}$, but we rely on the libraries that have efficient parallelized backends. The exact inverse of the full covariance matrix still scales as $\mathcal{O}(N_f^3N^3_p)$. Then we need to precompute ${}^{\ell_2\ell_1}(C^{-1})^{m_2m_1}_{\alpha\beta}$. The most efficient way to calculate it scales as $\mathcal{O}(\ell_{max}^2 N_p^2 + \ell_{max}^4 N_p)\times\mathcal{O}(N^2_f)$. The rest is the loop over the Fisher matrix indices with the trace over $m$ - $\mathcal{O}(N_f^2\ell_{max}^2)\times\mathcal{O}(\ell^2_{max})$ operations. We use $\texttt{numba}$ \cite{lam2015numba} for this part to handle compilation, parallelization, and customized logic to utilize all the symmetries. Hence, the total complexity is dominated by the $\mathcal{O}(\ell_{max}^2 N_p^2 + \ell_{max}^4 N_p)\times\mathcal{O}(N^2_f)$ term. As previously discussed, when the covariance is block diagonal, $N^2_f$ multiple is replaced by the number of fields in the largest block. We use $\texttt{fbpca}$ library to build the eigenvectors of the binned basis. In some places, we find it useful to use $\texttt{opt\_einsum}$ \cite{Smith2018} to evaluate contractions of multidimensional arrays.

\section{Validation on Simulations and Comparison with Pseudo-$C_\ell$}
\label{sec:validation}

In this section we validate our estimator on simulations, compare it with pseudo-C$_\ell$, and perform iterative parameter inference.

\subsection{Code validation: Unbiasedness and Mode-deprojection}

We validate our code using simple power-law power spectra and a realistic experimental mask.

\paragraph{Fiducial powers.}To evaluate the performance of our multi-field QML estimator, we simulated a set of 4 correlated Gaussian fields $a_1, a_2, b_1, b_2$, which follows an inverse-square power law fiducial angular power spectrum:
\begin{align}\label{eq:fiducial auto power}
    S^{a_1, a_1}_{\ell} &= S^{a_2, a_2}_{\ell} =  \ell^{-2}\, , \\
    S^{b_1, b_1}_{\ell} &= S^{b_2, b_2}_{\ell} = 0.8\,\ell^{-2}\, . \nonumber
\end{align}
To differentiate the observed autopower from the underlying signal, a uniform noise power spectrum $N_\ell = 10^{-4}$ is added to the auto-powers (we are signal dominated on all scales in this example), such that:
\begin{align}\label{eq:fiducial auto power w/ noise}
    C^{a_*, a_*}_{\ell} &= S^{a_*, a_*}_{\ell} + N_{\ell} \, , \\
    C^{b_*, b_*}_{\ell} &= S^{b_*, b_*}_{\ell} + N_{\ell} \, . \nonumber
\end{align}
where the $*$ symbol denotes any one of the two indexes (1, 2). It is noted that only $C^{a_*, a_*}_{\ell}, C^{b_*, b_*}_{\ell}$ are used in the simulation. We also consider a simple cross-correlation structure as follows:
\begin{align}\label{eq:fiducial cross power}
    C^{a_1, a_2}_{\ell} &= 0.5\,S^{a_1, a_1}_{\ell} \, , \\ 
    C^{b_1, b_2}_{\ell} &= 0.3\,S^{b_1, b_1}_{\ell} \, , \nonumber\\
    C^{a_*, b_*}_{\ell} &= A\times\sqrt{S^{a_1, a_1}_{\ell}\times S^{b_1, b_1}_{\ell}} \, , \nonumber
\end{align}
The cross-power amplitude $A$ is set to $0.4$. The maps are simulated on $N_{\text{side}} = 16$ resolution in \texttt{Healpix} format. The is achieved by the \texttt{hp.synalm} function available in the \texttt{Healpy} package~\cite{Zonca2019:Healpix1, 2005ApJ:Healpix2}. In the present work we use $N_{\text{side}} = 16$ to illustrate the approach, but higher $N_{\text{side}}$ are tractable and in particular we use $N_{\text{side}} = 32$ in our companion kSZ analysis paper.

\begin{figure}[h!]
    \centering
    \includegraphics[width=0.5\columnwidth]{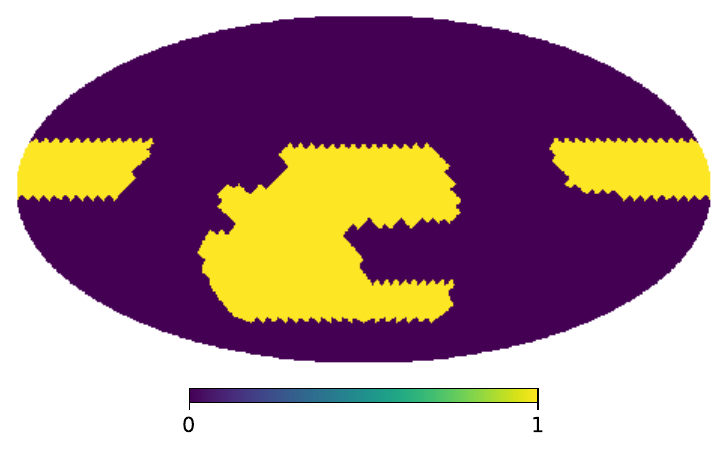}
    \caption{An $N_{\text{side}}=16$ binary mask downgraded from $N_{\text{side}}=32$ and applied to the kSZ analysis in~\cite{Lai2025kszpaper}, the unmasked pixels have a value of 1.}
    \label{fig:mask}
\end{figure}

\paragraph{Choice of mask.} To include the partial sky scenario analogous to the realistic survey geometry, we adopt a binary mask from~\cite{Lai2025kszpaper} that accounts for an overlapping region between the ACT DR6 Temperature map~\cite{DR6} and DESI LRG survey. The mask is obtained in $N_{side} = 32$ and then downgraded to $N_{side} = 16$ to match our resolution. Fig.~\ref{fig:mask} shows the mask that was used in this setup. The number of effective pixels reduces from $3072$ for the full-sky simulation to $766$ pixels for the masked sky.

\paragraph{Mode removal.} We investigate the robustness of the mode deprojection described in Sec.~\ref{subsec:deproj_ell} by removing the QML estimations below $\ell = 5$ through three separate methods.
\begin{enumerate}
\item Calculate the complete Fisher matrix $F_{\ell\ell'}$ and the estimator $Q_\ell$, then remove the rows and columns of the Fisher matrix corresponding to the unwanted modes. While this is the most direct approach, there is a big problem with it. Since the pixel-space inverse covariance matrix still involves unwanted fiducial modes $l < l_{\rm min}$ that are subjected to systematics, the power spectrum estimator, by construction, remains sensitive to those contaminated modes even if the couplings in the Fisher matrix are removed.

    \item Follow the prescription in~\cite{Tegmark_1997}, which is further elaborated and generalized to the multi-field scenario in Sec.~\ref{subsec:deproj_ell}. This is achieved by the construction of a pseudo-inverse covariance that spans a subspace orthogonal to the unwanted modes. This method is exact in a way that it removes the contribution from unwanted modes from the pixel-space covariance level. 

    \item Finally, we scale the unwanted modes by a factor of $10^3$ according to the prescription in~\cite{Tegmark_1997}. This approach utilizes that fact that the Fisher coupling scales roughly as $\sim\text{Tr}(C^{-2})$. Therefore, by scaling the unwanted modes in the fiducial power, one essentially suppresses the corresponding Fisher matrix element, which effectively 'decouples' the unwanted modes from the other.
\end{enumerate}
We comment on the different properties of these approaches below.

\begin{figure}[h!]
    \centering
    \includegraphics[width=0.85\columnwidth]{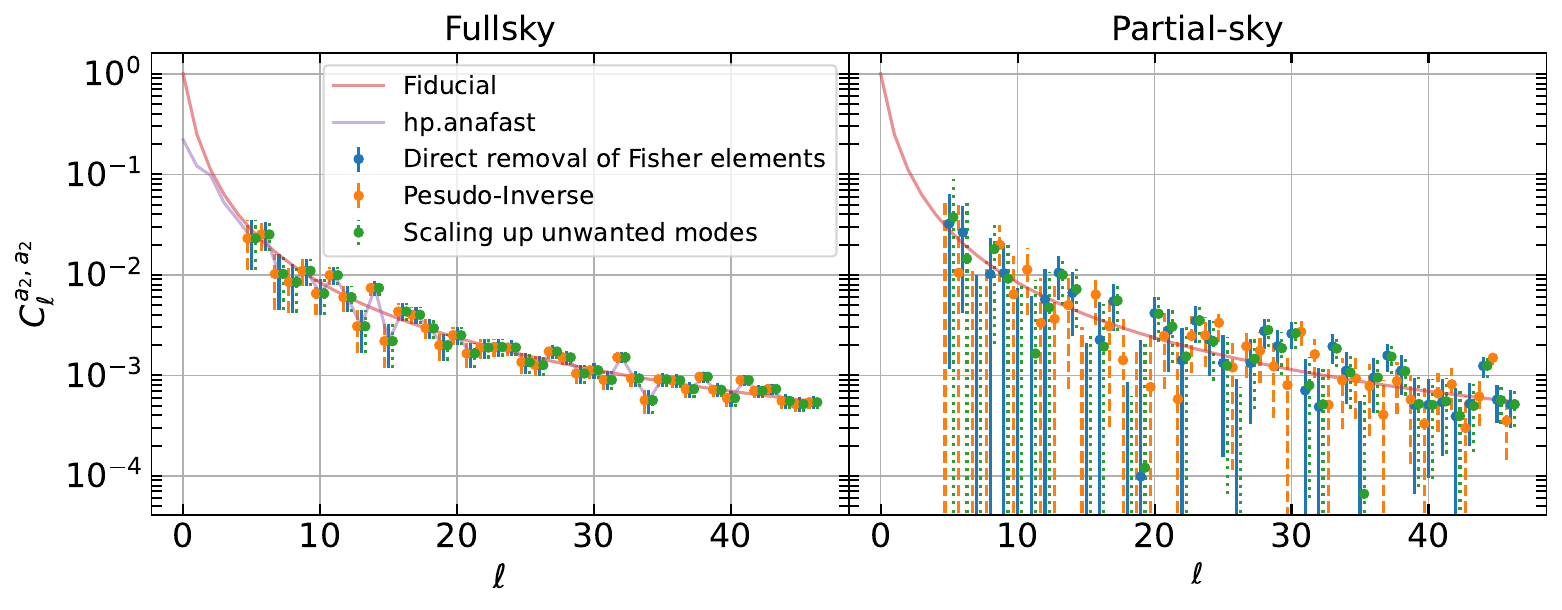}
    \caption{Comparison of mode-deprojection methods. Left: The full-sky QML estimates and the associated error bars for the $C^{a_2, a_2}_{\ell}$ power spectrum. The red curve shows the fiducial power spectrum given in Eq.~\ref{eq:fiducial auto power w/ noise} and the purple curve represents the power spectrum estimation from the fullsky Gaussian simulation using \texttt{hp.anafast}. Data points with solid, dashed and dotted error bars give the three different mode deprojection methods, with 'no treatment' stands for simple column/row removal (method 1), 'pinv' stands for pesudo-inverse covariance construction (method 2) and 'scale' for scaling unwanted modes (method 3). Right: Same as left panel, but with the mask shown in Fig.~\ref{fig:mask} applied to the same set of simulated maps. Green and orange data points are shifted by $\pm0.3$ along the x-axis to avoid overlapping.}
    \label{fig:QML_fullsky_partialsky}
\end{figure}

\paragraph{Comparing full sky to partial sky.} Fig.~\ref{fig:QML_fullsky_partialsky} compares the QML estimates from $\ell = 5$ to $\ell = 47$ for the three different mode-deprojection methods. One can observe that in the full-sky limit (left plot), three approaches give the same results. This is because in the full-sky limit, the modes are completely decoupled due to statistical isotropy, leading to a diagonal Fisher matrix. Therefore, the three mode-removal treatments are equivalent in this regime due to zero coupling between modes. Furthermore, it can be seen that all QML estimates converge to the naive fullsky power spectrum estimates obtained from the \texttt{hp.anafast} (purple curve), given by $\hat{C_\ell} = \frac{1}{2\ell+1}\sum_{m=-\ell}^{\ell}|a_{\ell m}|^2$, showing that our multi-field QML pipeline is optimal in the full-sky limit.

The three mode deprojection methods start to differ from each other in the cut-sky regime. With mode coupling introduced by the mask, the off-diagonal of the Fisher matrix $F_{\ell\ell'}$ is no longer zero. It can be seen that both the power spectrum estimates and the error bars are different for all three methods at $\ell < 30$. In particular, the data points obtained from the pseudo-inverse approach and the mode-scaling approach are similar to each other, revealing mode-scaling approach as a good approximation to the exact pseudo-inverse method. All three methods converge nicely at $\ell > 30$. The transition around $\ell = 30$ sets a characteristic scale at which mode-coupling becomes negligible.

\begin{figure}[h!]
    \centering
    \includegraphics[width=0.5\columnwidth]{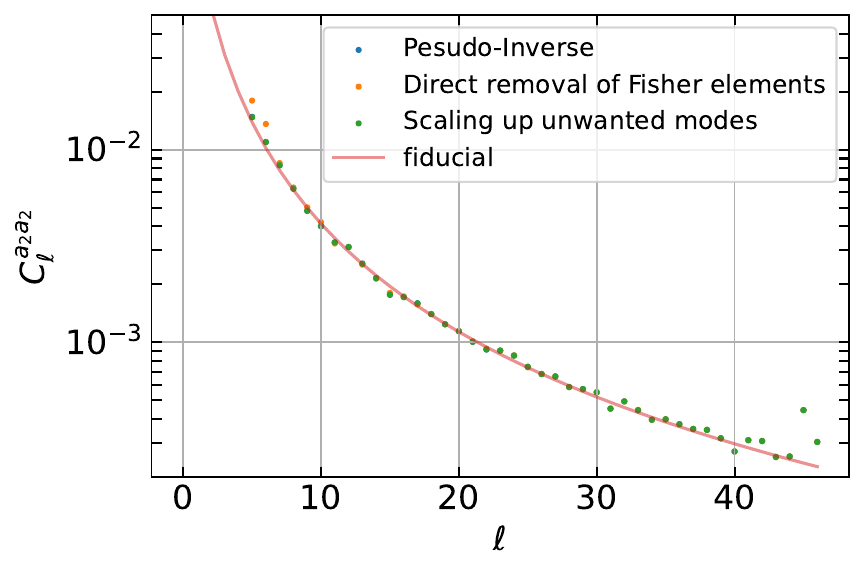}
    \caption{Comparison of (un-)biasedness of mode-deprojection methods. The QML estimates of the $C^{a_2, a_2}_{\ell}$ power spectrum averaged from 5000 masked Gaussian realizations. Each realization is processed with the multi-field QML pipeline and modes below $\ell = 5$ are removed according to the three prescriptions in main text. The fiducial power spectrum is also plotted for reference.
}
    \label{fig:QML_partialsky_biastest}
\end{figure}

\paragraph{(Un-)biasedness of different mode deprojection methods.} While Fig.~\ref{fig:QML_fullsky_partialsky} illustrates that different mode deprojection treatments can yield different results in the partial-sky regime, it does not provide evidence that the pseudo-inverse method is more appropriate. A good test to check whether the three methods are unbiased. We generate 5000 Gaussian simulations with mask applied and process them with the three different mode removal technique. The QML outputs are then averaged for each approach. We plot the result in Fig.~\ref{fig:QML_partialsky_biastest}. One can see both the pseudo inverse and mode scaling nicely converge to the fiducial power, while simply removing unwanted columns and rows from the Fisher matrix would lead to a bias around the boundary ($\ell = 5$) where the cut is applied, eventually giving QML estimates that overpredicts the power spectrum. There are also biases around the smallest mode $\ell =47$ for all three approaches, which is due to the low pixel resolution of the map.

\paragraph{Sub-optimal estimates due to wrong fiducial power.} Another feature of the QML method is that the power spectrum estimates are unbiased by construction regardless of the choice of fiducial power. This property guarantees that the QML always gives correct estimates to the full-sky power spectrum even if the fiducial power poorly explains the observation. Instead, a fiducial power that is different from the true power leads to a suboptimal estimator i.e., slightly larger error bars. We give a proof in App.~\ref{app:mismod_var} on how mismodeling fiducial power leads to suboptimal estimates. To illustrate it, we change the input fiducial power by modifying the signal $S^{aa}_{\ell}$ from $\ell^{-2}$ to $\ell^{-3}$ and Noise from $10^{-4}$ to $10^{-2}$

\begin{figure}[h!]
    \centering
    \includegraphics[width=0.6\columnwidth]{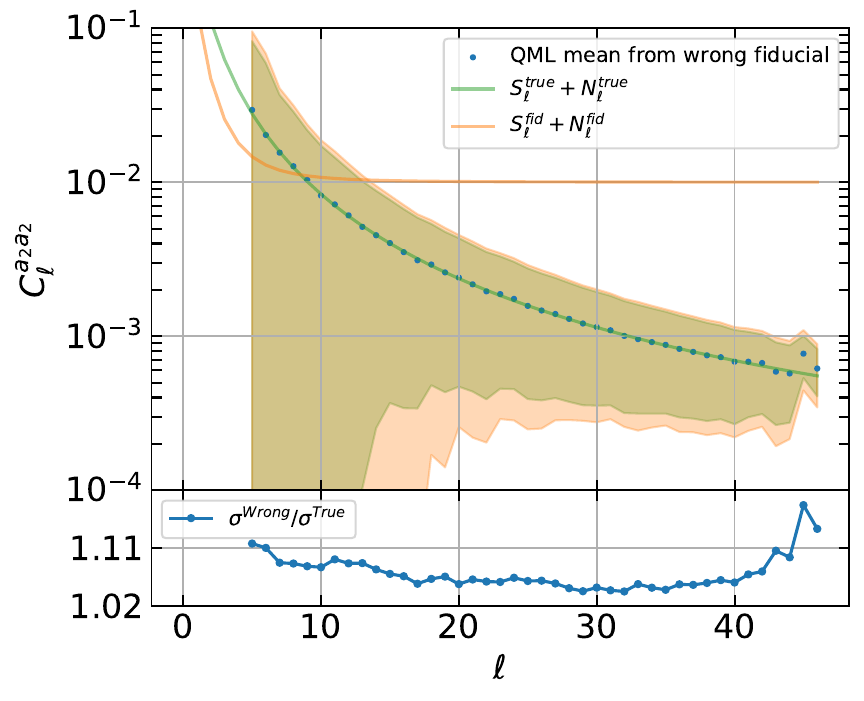}
    \caption{Influence of the fiducial power on the estimator optimality. The average of 5000 QML estimates, each realization is simulated from the true fiducial power spectrum given by green curve while evaluated with a modified covariance matrix that is computed from the orange curve. The green and orange shaded region highlight the 1-sigma uncertainty obtained from the 5000 simulations. The ratio of the 1-sigma uncertainty between Wrong and True fiducial power is plotted in the bottom, and it is consistently above 1.}
    \label{fig:QML_partialsky_wrongfiducial}
\end{figure}

We generate 5000 masked realization from the power spectrum without any modification and evaluate the QML output with the modified fiducial power spectrum. The blue data points in Fig.~\ref{fig:QML_partialsky_wrongfiducial} show the average of the QML estimates over the 5000 simulations. The fact that they fall on the true fiducial power shows that a mismodeled fiducial power still gives an unbiased estimation from the QML pipeline. The 1-sigma error of the simulation is highlighted as the orange region, which is larger than the optimal error bars computed analytically from the true fiducial power (green shaded region). 

As a result, constructing the QML pipeline with a wrong fiducial power leads to unbiased, yet suboptimal estimates when compared to the one that better describes the data. In this respect, the pseudo-$C_{\ell}$ formalism can be understood as a special case of mismodeling the fiducial covariance to be an identity matrix, as shown in~\cite{Terawaki:2025:weak_lensing}.

\subsection{Comparison with pseudo-C$_\ell$}

We now compare our estimator with the traditional pseudo-C$_\ell$ method that is widely used for power spectrum estimation on the sphere. For the pseudo-C$_\ell$ method, the covariance matrix needs to be determined by Monte Carlo, while in our QML case we calculate it exactly in a dense covariance matrix. Therefore, the pseudo-C$_\ell$ analysis in total is often slower than our approach, for the low angular resolution where the QML is tractable. Of course, the pseudo-C$_\ell$ are tractable to far higher angular resolution.

\begin{figure}[h!]
    \centering
    \includegraphics[width=0.5\columnwidth]{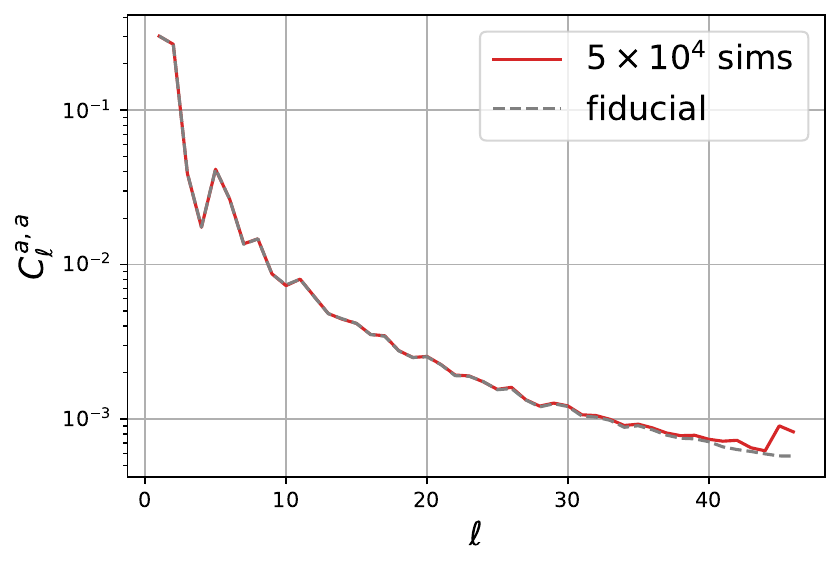}
    \caption{Test of the pseudo-$C_{\ell}$ pipeline. Pseudo-$C_{\ell}$ estimates on the autopower $C^{a,a}_{\ell}$ from averaging $10^5$ pseudo-$C_{\ell}$ outputs. The pseudo-$C_{\ell}$ power spectrum is calculated using \texttt{NaMaster} without mask apodization~\cite{Alonso:2018namaster}. The red curve shows the fiducial power spectrum convolved with the bandpower window function. There is residual bias beyond $\ell = 46$. The spikiness of the fiducial power is due to the window function.}
    \label{fig:pcl_convergence}
\end{figure}

\paragraph{Comparison to pseudo-$C_{\ell}$: Single Field.} We compare the performance of the QML estimator with the more conventional pseudo-$C_{\ell}$ approach in the single field case generated from the autopower spectrum $C^{a_*, a_*}_{\ell}$ in Eq.~\ref{eq:fiducial auto power w/ noise}. The pseudo-$C_{\ell}$ pipeline is implemented using NaMaster~\cite{Alonso:2018namaster}. We use the mask introduced in Fig.~\ref{fig:mask} as the input and construct bandpowers from $\ell = 1$ to $\ell = 46$ on a $\Delta\ell = 1$ basis. It is known that pseudo-$C_{\ell}$ estimates could be biased when the geometry of the mask is too complex, or when mask is not appropriately apodized. To test the pseudo-$C_{\ell}$ in an equivalent setting to the QML, we inject $10^5$ realizations to the pseudo-$C_{\ell}$ and plot the average power spectrum in Fig.~\ref{fig:pcl_convergence}. Modes beyond $\ell = 46$ are identified to be unreliable (at the border of our angular resolution). No apodization is needed to achieve unbiasedness at this low angular resolution.

\begin{figure}[h!]
    \centering
    \includegraphics[width=0.9\columnwidth]{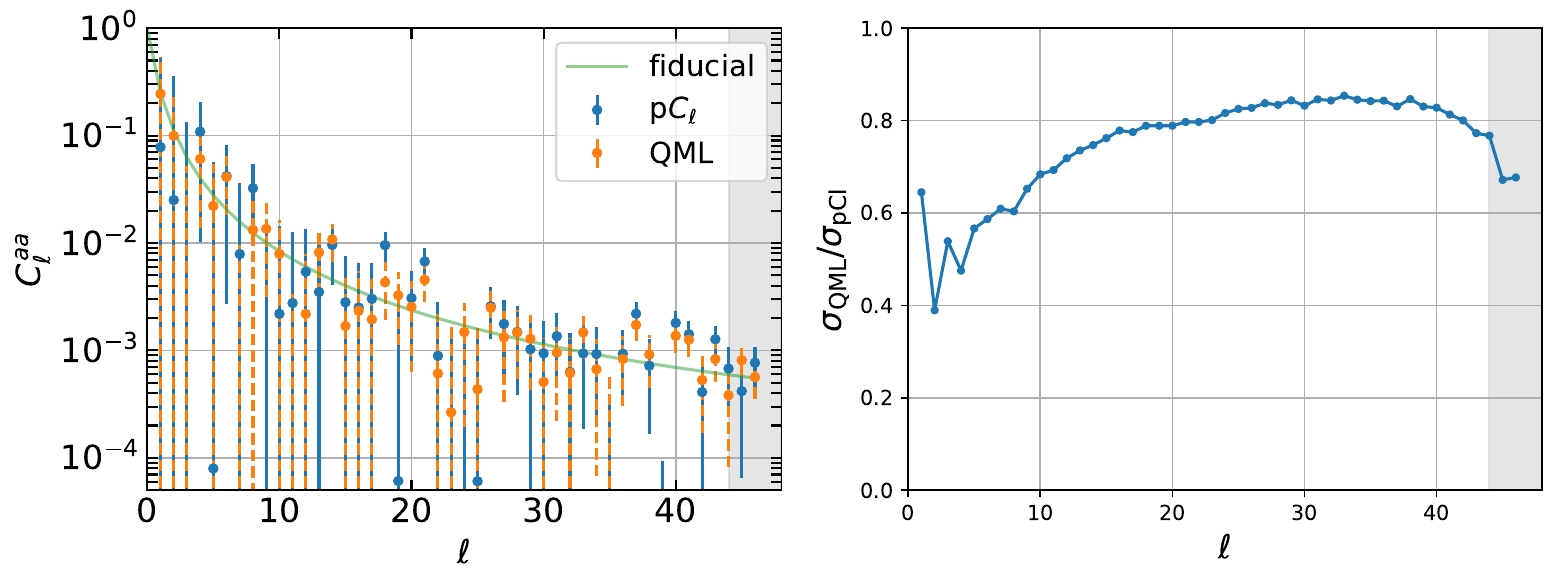}
    \caption{Comparison of pseudo-$C_{\ell}$ and QML error bars. Left: A comparison between pseudo-$C_{\ell}$ estimate and QML estimate of the autopower $C^{a, a}_{\ell}$ for the same input masked maps. The pseudo-$C_{\ell}$ power spectrum is evaluated with the \texttt{NaMaster} pipeline starting at $\ell = 1$. $\ell = 0$ has been deprojected with a pseudo-inverse covariance for QML output. The grey shaded region highlights the modes that fail to converge for both methods. Right: The ratio of the 1-sigma uncertainty between the QML estimates and the pseudo-$C_{\ell}$ output. QML yields the greatest improvement on large scales and gradually approaching to a constant ratio as $\ell$ increases.}
    \label{fig:QML_vs_PCL_1field}
\end{figure}

The left panel of Fig.~\ref{fig:QML_vs_PCL_1field} visualizes the QML and pseudo-$C_{\ell}$ power spectrum with their 1-sigma error bar. Note that while the pseudo-$C_{\ell}$ estimates are consistent with the fiducial power (green curve) within 1-sigma, they are more scattered compared to the QML estimates. The better quality of QML estimates is also illustrated in terms of the ratio of the error bars between the two methods, which increases from around 0.5 for large scale to 0.85 for small scale. For the single-field case, this ratio can be directly translated to the net improvement from switching the pseudo-$C_{\ell}$ estimator to a pixel-space QML estimator.

\paragraph{Comparison to pseudo-$C_{\ell}$: Two fields.}
For a setup with more than one input fields, the QML outputs the autopower spectrum for all fields and crosspower spectrum for any pair of fields. Parameter estimation can often benefit from the crosspowers, since they are more robust to systematics and calibration than the autopowers. Pseudo-$C_{\ell}$ consider only the information from the pair of associated field when constructing estimators for crosspowers. On the other hand, a multi-field QML pipeline constructs crosspower estimators by accounting the correlation of all fields, hence provides extra constraining power to the variance on top of its pixel-space nature.

We found that the QML outputs benefit the most from cross-correlation when the autopowers of fields are in a deep signal-dominant regime and differ a lot from each other. To illustrate this, we consider a simplified setup of one type $a$ field and one type $b$ field, where the fiducial power of $b$ field $S^{b,b}$, noise level $N_{\ell}$ and cross-correlation coefficient $A$ are modified as follows:
\begin{align}
    S^{a,a}_{\ell} = \ell^{-0.5}&,\,\, S^{b,b}_{\ell} = \ell^{-2}\, ,\\
    N_{\ell} =& 10^{-4} \, , \\
    A =& 0.9 \, ,
\end{align}
such that $C^{a,a}_{\ell} \approx S^{a,a}_{\ell} \gg S^{b,b}_{\ell} \approx C^{b,b}_{\ell}$. We consider two configurations of input fiducial power here, one with a dense pixel-space covariance matrix that incorporates autopowers and the crosspowers. The fiducial crosspower, $C^{a,b}_{fid}$, is set to zero for another configuration. This procedure follows the prescription in Sec.~\ref{subsec:deprojx} and helps the crosspower estimator to avoid contribution from autopowers, which could be less reliable due to systematics.

\begin{figure}[h!]
    \centering
    \includegraphics[width=0.9\columnwidth]{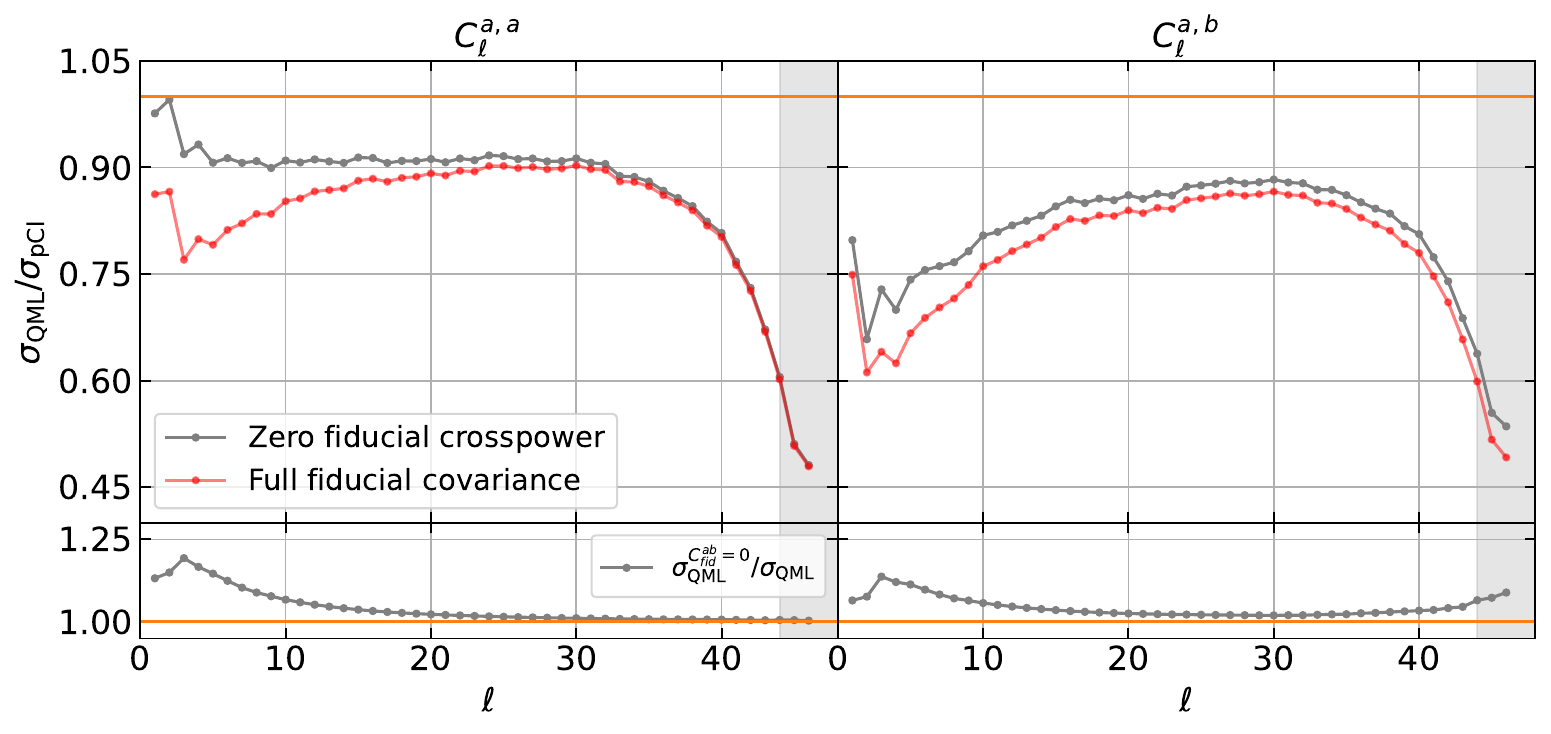}
    \caption{Influence of fiducial cross-powers on estimator error bars. Left: The ratio of QML error bar to the pseudo-$C_{\ell}$ error bars for the autopower $C^{a,a}_{\ell}$. Pseudo-$C_{\ell}$ error bars are evaluated from $5\times10^4$ simulations. The QML error bars represented in the gray curves are calculated using a pixel-space fiducial covariance matrix without crosspower contribution, while red curve uses the full covariance matrix. The bottom panel shows the ratio of the gray curve to the red curves. Right: Same as the left panel but for the crosspower $C^{a,b}_{\ell}$. An orange line at 1 is included in every plot to help identify that the ratio is consistently above/below 1.}
    \label{fig:QML_vs_PCL_2field}
\end{figure}

The left panel of Fig.~\ref{fig:QML_vs_PCL_2field} displays the ratio of the QML error bars to that of the pseudo-$C_{\ell}$ estimation for autopower $C^{aa}_{\ell}$ and cross power $C^{ab}_{\ell}$ on the right panel. The gray curve illustrates the configuration in which the fiducial crosspower is zeroed out in the QML pipeline, while the red curve utilizes the dense covariance matrix with the fiducial crosspower included. As expected, both the red and gray curves are consistently below 1, demonstrating the optimal nature of QML compared to pseudo-$C_{\ell}$. In addition, since zeroing out the fiducial crosspower can be understood as a special case of mismodeling the underlying fiducial power, the QML error bar should be larger than the one computed from the full covariance matrix. This is depicted by the ratio between the gray curve to the red curve, which is given by the bottom panel of each plot. 

The right plot in Fig.~\ref{fig:QML_vs_PCL_2field} answers an interesting question: How much of the improvement of a cross-power measurement with the QML over the pseudo-$C_{\ell}$ comes from the better pixel-wise weighting, and how much comes from the additional inclusion of auto-powers in the estimator (which is often undesirable because these are affected by systematics and calibration issues)? We see that for our setup, the improvement comes primarily from the better weighting.

\subsection{Band-powering of the Fisher matrix} 
\begin{figure}[h!]
    \centering
    \includegraphics[width=0.5\columnwidth]{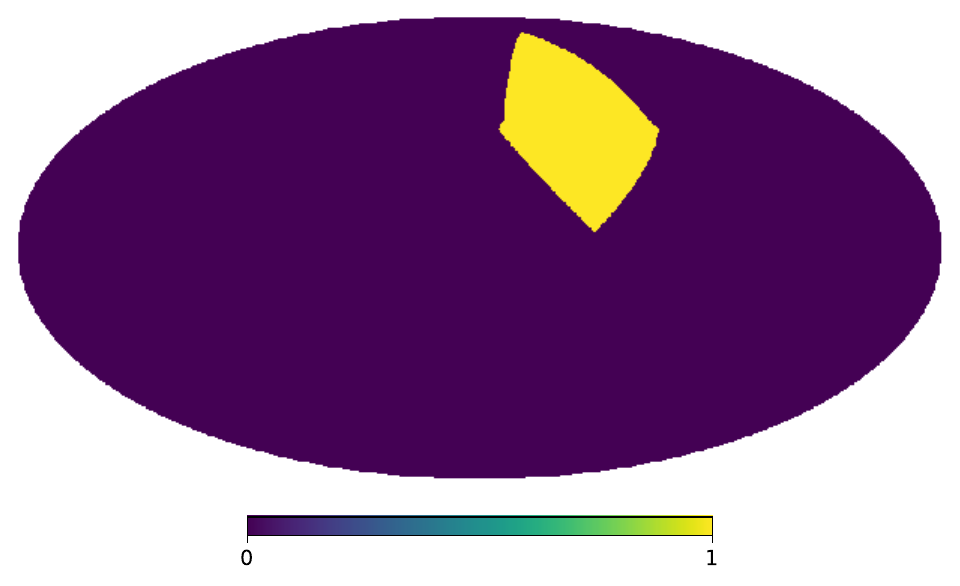}
    \caption{Small sky fraction mask. An $N_{\text{side}}=16$ binary mask covering a connected patch in the sky with sky fraction is 0.055, the restricted minimum angular size of the unmasked region leads to a singular Fisher matrix.}
    \label{fig:bandpower_mask}
\end{figure}

\begin{figure}[h!]
    \centering
    \includegraphics[width=0.9\columnwidth]{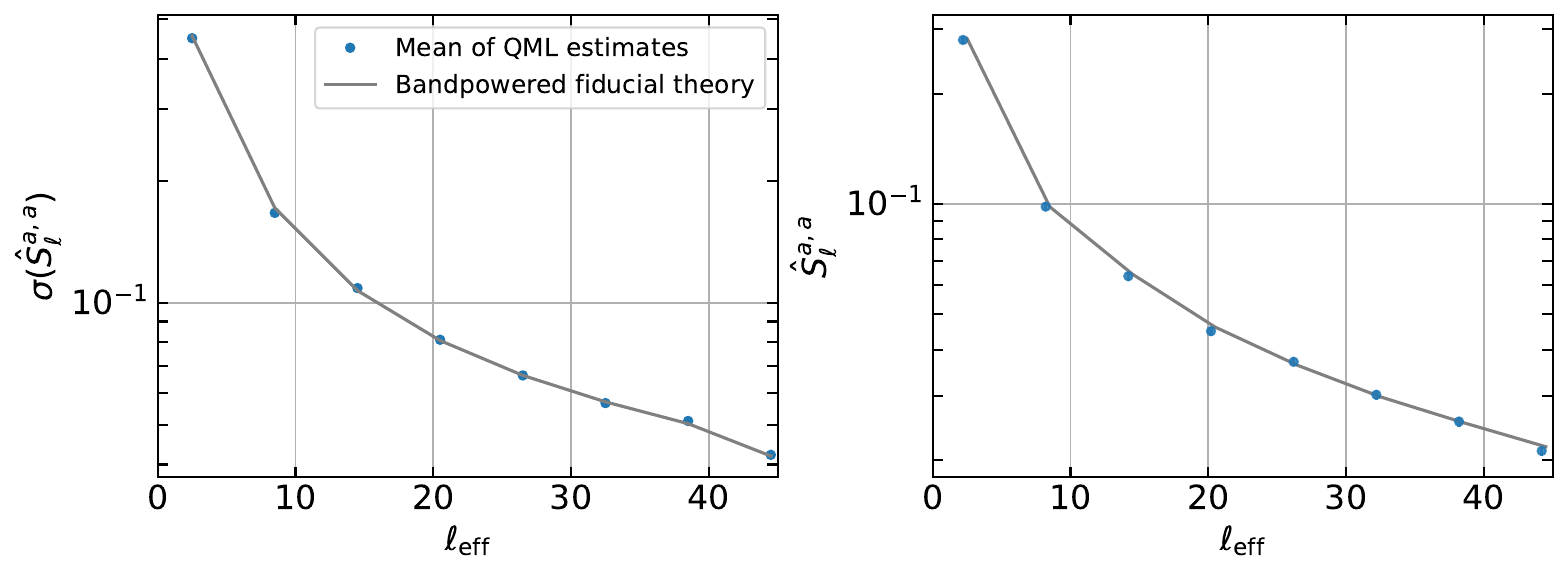}
    \caption{Bandpowered analysis. The bandpowering is performed by uniformly binning every 6 modes between $\ell = 0$ to $\ell = 47$. Left: The error bars of the bandpowered QML estimates $\sigma(\hat{S}^{a,a}_{\ell})$. The grey curve shows the analytical calculation from the diagonal elements of the bandpowered Fisher matrix defined in Eq.~\ref{eq:Bandpowered Fisher}. The blue data points are computed from taking the covariance of 4000 simulations, showing agreement with the theory. Right: Similar to the left plot but for the power spectrum estimates $\hat{S}^{a,a}_{\ell}$, the error bars from averaging over 4000 Gaussian simulations are included in the plot but too small to be seen.
    }
    \label{fig:QML_bpw}
\end{figure}

Following the procedure in Sec.~\ref{subsec:Bandpowered estimator}, we also demonstrate how bandpowering can be performed to deal with the non-invertability of the Fisher matrix $F_{\ell,\ell'}$ in the case of a severely limited sky region. We use the sky mask in Fig.~\ref{fig:bandpower_mask} with a sky fraction of around 0.055 for this test. We also adopt a single-field configuration that follows the fiducial power setup for $S^{a,a}_{\ell}$ in Eq.~\ref{eq:fiducial auto power w/ noise} with a uniform noise power spectrum of $N^{aa}_{\ell} = 10^{-1}$. While $F_{\ell, \ell'}$ is well-defined, it is poorly-conditioned for inversion. We therefore employ a bandpowering matrix $P_{b\ell}$ to the uncoupled estimators $\hat{y}_{\ell}$ that bins every 6 multipoles from $\ell = 0$ to $\ell = 47$ with a uniform window function. We show the analytical error bars in the left panel of Fig.~\ref{fig:QML_bpw} as the grey curve; we also show bandpowered fiducial theory in the right panel of the same figure. The mean of the QML estimates, computed from 4000 Gaussian simulations, are illustrated as the blue data points on both panels. Good convergence can be observed as the data points align with the fiducial theory.

\subsection{Iterative parameter inference}

Finally, we discuss the use of the QML for power-spectrum-based parameter inference. 

\paragraph{QML as a parameter inference.}The fact that the pixel covariance matrix implicitly depends on the fiducial power spectrum relates the optimal QML estimator to the inference of power spectrum modeling parameters. If the modeling parameters are inferred from the initial QML estimates, the initial fiducial power leads to suboptimal error bars (see the orange shaded region in Fig.~\ref{fig:QML_partialsky_wrongfiducial}, for instance). As mentioned in~\cite{Bilbao_Ahedo_2021}, the mismatch between an initial guess on the fiducial power and the one that generates the observed maps can be addressed through an iterative process, where the fiducial power for the next QML estimator is updated from the output of the previous QML. In this section, instead of using a smoothened QML power spectrum as the new input, we explore the convergence to optimal estimate within the framework of a given model, such that the best-fit modeling parameters are inferred and used to compute the new fiducial power.

\begin{figure}[h!]
    \centering
    \includegraphics[width=0.9\columnwidth]{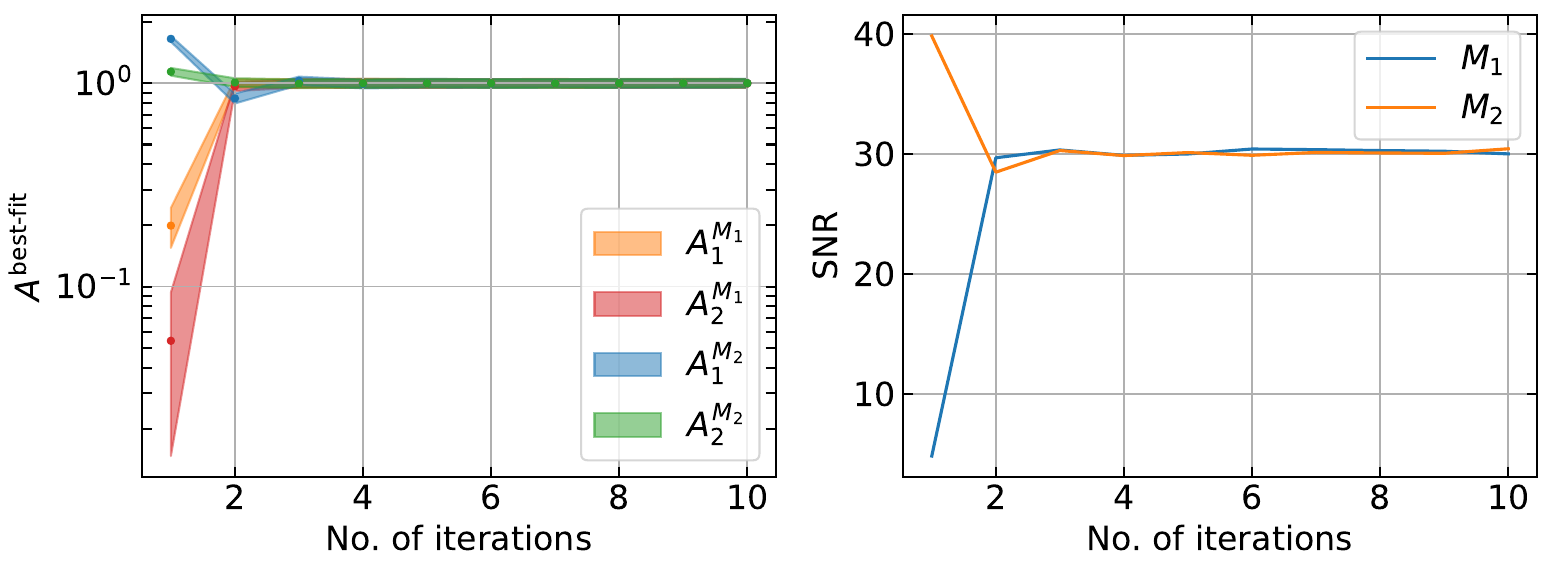}
    \caption{Iterative QML application. Left: The best-fit modeling parameter $A^{\text{best-fit}}_a, A^{\text{best-fit}}_b$ from taking the average of the MCMC sampling chain for the two mismodeled fiducial powers $M_1, M_2$ in 10 iterations. The shaded regions denotes the 1-sigma uncertainties, which is also calculated from the MCMC chains. All the modeling parameters quickly approach to the fiducial values $A_a = A_b = 1$ that are used to generate the observations. Right: The SNR computed according to Eq.~\ref{eq:SNR definition} for the two models. Depending on the exact choice of initial fiducial power, the SNR can over or undershoots the converging value.}
    \label{fig:QML_iteration}
\end{figure}

We generate a set of observed maps according to the fiducial powers in Eq.~\ref{eq:fiducial auto power w/ noise},~\ref{eq:fiducial cross power}. However, we construct the covariance matrix with modified auto-powers:
\begin{align}\label{eq:mismodelling parameters}
    \tilde{S}^{a_*, a_*}_{\ell} &=  A_a S^{a_*, a_*}_{\ell}\, , \nonumber \\
    \tilde{S}^{b_*, b_*}_{\ell} &=  A_b S^{b_*, b_*}_{\ell}\, , \nonumber \\
    \tilde{C}^{a_*, b_*}_{\ell} &= A\times\sqrt{\tilde{S}^{a_1, a_1}_{\ell}\times \tilde{S}^{b_1, b_1}_{\ell}} \, , \nonumber
\end{align}
Note that the two parameters $A_a$ and $A_b$ are included to introduce mismodeling into the covariance matrix so that the simulated maps have a fiducial setup of $A_a = 1, A_b = 1$. We consider two models $M_1: (A_a = 5, A_b = 20)$ and $M_2: (A_a = 0.7, A_b = 0.9)$. For each of the models, we construct a QML pipeline and inject the simulated maps to obtain power spectrum estimates. We assume the power spectrum estimates follow a Gaussian likelihood: 
\begin{equation}
\label{eq:log likelihood function}
    -2\ln\mathcal{L}(A_a, A_b|\hat{\tilde{C}}_{\ell}) = (\hat{\tilde{C}}_{\ell} - \tilde{C}_{\ell}(A_a, A_b))^T \tilde{F}_{\ell\ell'}(\hat{\tilde{C}}_{\ell'} - \tilde{C}_{\ell'}(A_a, A_b)) \, ,
\end{equation}
the quantity $\hat{\tilde{C}}_{\ell}$ should be understood as a vector of all QML estimates. $\tilde{F}_{\ell\ell'}$ is the analytic Fisher matrix from the QML pipeline, the tilde notation denotes a quantity that is derived from the mismodeled fiducial power. For this test case, we restrict ourselves to the QML estimates between $5 \leq \ell \leq 45$, where a Gaussian likelihood provides a good approximation. However, it is noted that the Wishart distribution should instead be used for analysis that includes modes with $\ell < 5$. To demonstrate the flexibility to extend to more complex and nonlinear models, we perform a Markov Chain Monte-Carlo (MCMC) sampling using the package \texttt{emcee} to determine the best-fit parameters $A_1, A_2$. The best-fit parameters from each iteration are then used to update the fiducial model in the following fashion:
\begin{align}\label{eq:fiducial model update}
    A^{i+1}_a = A^{i, \text{best-fit}}_a\times A^{i}_a \, , \nonumber \\
    A^{i+1}_b = A^{i, \text{best-fit}}_b\times A^{i}_b \, ,
\end{align}
where $A^{i, \text{best-fit}}_a$ is the best-fit parameter in the i-th iteration and $A^{i}_a$ is the modeling parameter that is used to construct the QML covariance matrix for the i-th iteration. The left panel of Fig.~\ref{fig:QML_iteration} shows the best-fit parameters computed from the mean of MCMC chains for 10 iterations, note that the modeling parameters for both models $M_1$ and $M_2$ already converge to the fiducial value $A_1 = A_2 = 1$ in the second iteration. The plot on the right displays the signal-to-noise ratio (SNR), defined as:
\begin{align}\label{eq:SNR definition}
    \text{SNR}^i &= \sqrt{(\mathbf{A}^{i})^T \text{Cov}^{-1}(\mathbf{A}^i)\mathbf{A}^i} \, ,
\end{align}
here $\mathbf{A}^{i} = (A^{i, \text{best-fit}}_a,  A^{i, \text{best-fit}}_b)$ is the flattened best-fit modeling parameter and $\text{Cov}^{-1}(\mathbf{A}^i)$ is the 2$\times$2 covariance matrix evaluated from the MCMC chains. While the SNR converges to the same value (30) and at the same rate as the modeling parameter for both models, one has to be careful about evaluating the SNR without any iterations, since it could be strongly biased by choice of fiducial power.

\paragraph{Different approaches in updating the fiducial power.} We recall that in~\cite{Bilbao_Ahedo_2021}, the fiducial power is updated using a smoothed QML output from the previous iteration. This method allows the QML pipeline to iterate for globally optimal errorbars that fit the observed data the best. Nevertheless, in more common circumstances where errorbars and SNR are quantified with respect to a given model with a certain number of free parameters, the iterative process in~\cite{Bilbao_Ahedo_2021} becomes less applicable. In our approach we combine QML estimation with parameter inference, either through MCMC sampling or likelihood maximization over the model space, such that the fiducial powers can be improved within restricted degrees of freedom. As shown in Fig.~\ref{fig:QML_iteration}, a local optimality can still be achieved by updating the best-fit modeling parameters, hence one can quantify the optimal error bars and SNR for a given model.

\section{Conclusion}
\label{sec:conclusion}
In this work, we developed the python-based QML estimator code \texttt{QML-FAST} for power spectrum estimation of redshift-binned spherical data, based on the QML formalism in~\cite{Tegmark_1997, Tegmark_2001}. While the QML has been used in modern CMB data analysis for example in ~\cite{Bilbao_Ahedo_2021} or for weak-lensing data in~\cite{Maraio_2023}, our pipeline extends the QML estimator to a large number of correlated fields. We incorporated a series of tests to illustrate the unbiasedness and optimality of our pipeline. We also employed a pseudo-inverse covariance construction to deproject unwanted modes from the pixel-space covariance matrix. By combining the QML output with parameter inference, we showed that an iterative scheme can be applied to optimize the best-fit parameters and variances with respect to a fixed fiducial model. Various computational optimizations, including the use of highly efficient libraries like \texttt{opt\_einsum} and \texttt{numba}, sparse matrix multiplication, and change of basis have been introduced to our pipeline to reduce the complexity of the calculation and to make memory allocation more efficient. For example, to get the QML estimates for the 2-by-2 fields configuration on $N_{\text{side}} = 16$ resolution requires less than a minute on a 24-core CPU. A more complicated and physical scenario presented in the companion paper~\cite{Lai2025kszpaper}, which contains 40-by-40 fields configuration on $N_{\text{side}} = 32$, takes only about an hour. The code, together with documentation and associated tutorial notebooks, is made available in~\url{https://github.com/ykvasiuk/qmlfast}. Our code will allow the optimal analysis of binned photometric galaxy survey data from current and upcoming surveys on large scales, in particular, primordial non-Gaussianity search with scale-dependent bias and kSZ velocity reconstruction. Because it features a binned pixel-wise covariance matrix, our code could also be useful to examine and de-project large-scale systematics that affect such analyses.

\section*{Acknowledgements}
M.M., Y.K. and A.L acknowledge the support by the U.S. Department of Energy, Office of Science, Office of High Energy Physics under Award Number DE-SC0017647, the support by the National Science Foundation (NSF) under Grant Number 2307109 and 2509873 and
the Wisconsin Alumni Research Foundation (WARF). Y.K. is grateful for the hospitality of Perimeter Institute,  where a part of this work was done. KMS was supported by an NSERC Discovery Grant, by the Daniel Family Foundation, and by the Centre for the Universe at Perimeter Institute. Research at Perimeter Institute is supported in part by the Government of Canada through the Department of Innovation, Science and Economic Development Canada and by the Province of Ontario through the Ministry of Colleges and Universities.
\printbibliography
\appendix
\section{Variance due to mismodeling of the signal covariance}
\label{app:mismod_var}
Here, we show that the QML estimator, built with $\mathbb{C}_{fid}\neq\mathbb{C}_{true}$ is suboptimal. Let us \begin{itemize}
    \item assume we mismodel the signal covariance, but not the noise, so that the estimator is unbiased;
    \item absorb the noise term in the signal for brevity;
    \item denote with tildas the quantities derived from the mismodelled data covariance
\end{itemize} 
Then the following holds
\begin{equation}
        \langle \tilde{y}_A\rangle = \langle\pmb{\phi^T}\tilde{Q}_A\pmb{\phi}\rangle = \tilde{F}_{AB}c_B
\end{equation}
Without the mismodelling, we have
\begin{equation}
        \langle y_A\rangle = \langle\pmb{\phi^T}Q_A\pmb{\phi}\rangle = F_{AB}c_B
\end{equation}
Gaussian $\pmb{\phi}$ and symmetric $A,B$ satisfy
\begin{equation}
    \mathrm{cov}(\pmb{\phi^T}A\pmb{\phi}, \pmb{\phi^T}B\pmb{\phi}) = 2\texttt{Tr}[\mathbb{C}A\mathbb{C}B]
\end{equation}
Let's consider $\mathfrak{y}_A = \begin{pmatrix}
    y_A \\
    \tilde{y}_A
\end{pmatrix}$.
For the covariance $\mathrm{cov}(\mathfrak{y}_A,\mathfrak{y}_B)$, we can calculate:
\begin{itemize}
    \item $\mathrm{cov}(y_A,y_B) = F_{AB}$
    \item $\mathrm{cov}(y_A,\tilde{y}_B) = \frac{1}{2}\texttt{Tr}[\mathbb{C}\mathbb{C}^{-1}P_A\mathbb{C}^{-1}\mathbb{C}\tilde{\mathbb{C}}^{-1}P_B\tilde{\mathbb{C}}^{-1}] \equiv \tilde{F}_{AB}$
    \item $\mathrm{cov}(\tilde{y}_A,\tilde{y}_B) = \frac{1}{2}\texttt{Tr}[\mathbb{C}\tilde{\mathbb{C}}^{-1}P_A\tilde{\mathbb{C}}^{-1}\mathbb{C}\tilde{\mathbb{C}}^{-1}P_B\tilde{\mathbb{C}}^{-1}] \equiv G_{AB}$
\end{itemize}
So that
\begin{equation}
    \mathfrak{C}_{AB} = \mathrm{cov}(\mathfrak{y}_A,\mathfrak{y}_B) = \begin{pmatrix}
        F_{AB} & \tilde{F}_{AB} \\
        \tilde{F}_{AB} & G_{AB}
    \end{pmatrix}
\end{equation}
Because $\mathfrak{C}_{AB}$ is a covariance, it has to be at least positive semi-definite:
\begin{equation}
    \mathfrak{C}_{AB} 	\succeq 0
\end{equation}
(Statement "$A\succeq 0$" is equivalent to the statement "for any non-zero $x$, $x^TAx \geq 0$"). We assume that the inverses $\tilde{F}^{-1}$ and $F^{-1}$ exist. Then it follows that 
\begin{equation}
\label{eq:mism_var}
    \tilde{F}^{-1}G\tilde{F}^{-1} - F^{-1} \succeq 0
\end{equation}
It suffices to show that $v^T \big( \tilde{F}^{-1}G\tilde{F}^{-1} - F^{-1} \big) v \geq 0$ for any vector $v\ne 0$:
\begin{equation}
v^T \big( \tilde{F}^{-1}G\tilde{F}^{-1} - F^{-1} \big) v
  = \begin{pmatrix} -F^{-1}v & \tilde{F}^{-1}v \end{pmatrix}
  \underbrace{\begin{pmatrix}
      F & \tilde{F} \\ \tilde{F} & G
  \end{pmatrix}}_{=\mathfrak{C}_{AB}}
  \begin{pmatrix} -F^{-1}v \\ \tilde{F}^{-1}v \end{pmatrix}
  \geq 0  \hspace{1cm} \mbox{since } \mathfrak{C}_{AB} \succeq 0
\end{equation}
We can recognize the first term from the left in the equation \eqref{eq:mism_var} as the covariance matrix of the estimator $\tilde{\hat{c}}$.

\section{Pseudo-inverse in the case of $N$ fields}
\label{app:pi}
We are going to repeat the argument of \cite{Tegmark_1997} (Appendix C) first and then show how it extends to the general case. The original argument defines an orthogonal $N_p\times N_p$ matrix $R$ such that its first $\ell_{min}$ columns and rows are equal to the rows and columns of $Z$. In this basis, $\pi$ takes block-diagonal form:
\begin{equation}
    R\pi R^T = R[1 - ZZ^T]R^T = \begin{pmatrix}
        0 & 0 \\
        0 & 1
    \end{pmatrix}
\end{equation}
So does the deprojected covariance:
\begin{equation}
    R\pi C \pi^TR^T = \begin{pmatrix}
        0 & 0 \\
        0 & C_*
    \end{pmatrix}
\end{equation}
And for the pseudo-inverse $M$ one has in the same basis
\begin{equation}
    RMR^T = \begin{pmatrix}
        0 & 0 \\
        0 & 1
    \end{pmatrix} \left[\begin{pmatrix}
        0 & 0 \\
        0 & C_*
    \end{pmatrix}+ \begin{pmatrix}
        \eta1 & 0 \\
        0 & 0
    \end{pmatrix}\right]^{-1}\begin{pmatrix}
        0 & 0 \\
        0 & 1
    \end{pmatrix} = \begin{pmatrix}
        0 & 0 \\
        0 & C_*^{-1}
    \end{pmatrix}
\end{equation}
To see why the same argument works in the case of $N$ fields, it is useful to recall the properties of the commutation matrix $S$. Let $A$ be a $(p\times q)$ matrix and $B$ - $(m\times n)$ matrix. Then one of the equivalent definitions of the $(ab \times ab)$ matrix $S^{(a,b)}$ is via the property
\begin{equation}
    S^{(m,p)}(A \otimes B)S^{(q,n)} = B \otimes A
\end{equation}
Commutation matrix $S$ is orthogonal: 
\begin{equation}
    (S^{(a,b)})^T = S^{(b,a)} = (S^{(a,b)})^{-1}
\end{equation}
Throughout this paper, we have naturally assumed the covariance of the block form with $N_f\times N_f$ blocks of the size $N_p\times N_p$. More formally, we define decomposition
\begin{equation}
    \mathbb{C} = \sum_{ab \in \mathcal{F}}\sum_{ij \in \mathcal{P}} c_{ab;ij}E^{ab}_{\mathcal{F}} \otimes E^{ij}_\mathcal{P}
\end{equation}
with $\mathcal{F}$ and $\mathcal{P}$ representing fields and pixels correspondingly. It follows then that the action of $S^{(N_f,N_p)}$ switches the order of basis elements:
\begin{equation}
    \mathbb{C'} = S\mathbb{C}S^T = \sum_{ab \in \mathcal{F}}\sum_{ij \in \mathcal{P}} c_{ij;ab}E^{ij}_{\mathcal{P}} \otimes E^{ab}_\mathcal{F}
\end{equation}
For the projection operator $\Pi$, one gets:
\begin{equation}
    \mathrm{S}\Pi\mathrm{S^T} = \mathrm{S}[\mathbbm{1}_{N_f}\otimes \pi]\mathrm{S^T} = \pi \otimes \mathbbm{1}_{N_f}
\end{equation}
Without loss of generality, let's consider a case of two fields. The deprojected covariance takes the following form:
\begin{equation}
    \mathrm{\Pi}\mathbb{C}\mathrm{\Pi}^T = \begin{pmatrix}
        \pi C_{11} \pi^T & \pi C_{12} \pi^T \\
        \pi^T C^T_{12} \pi & \pi C_{22} \pi^T
    \end{pmatrix}
\end{equation}
In the basis defined by matrix $\mathbb{R} = \mathbbm{1}_{N_f} \otimes R$, we have:
\begin{equation}
\mathbb{R}\mathrm{\Pi}\mathbb{C}\mathrm{\Pi}^T\mathbb{R}^T = \begin{pmatrix}
        \begin{pmatrix}
        0 & 0 \\
        0 & C^{11}_{*}
    \end{pmatrix} & \begin{pmatrix}
        0 & 0 \\
        0 & C^{12}_*
    \end{pmatrix} \\
        \begin{pmatrix}
        0 & 0 \\
        0 & C^{21}_*
    \end{pmatrix} & \begin{pmatrix}
        0 & 0 \\
        0 & C^{22}_*
    \end{pmatrix}
    \end{pmatrix}
\end{equation}
The action of the commutation matrix S then gives:

\begin{equation}
\mathrm{S}\mathbb{R}\mathrm{\Pi}\mathbb{C}\mathrm{\Pi}^T\mathbb{R}^T\mathrm{S^T} = \begin{pmatrix}
        \begin{pmatrix}
        0 & 0 \\
        0 & 0
    \end{pmatrix} & \begin{pmatrix}
        0 & 0 \\
        0 & 0
    \end{pmatrix} \\
        \begin{pmatrix}
        0 & 0 \\
        0 & 0
    \end{pmatrix} & \begin{pmatrix}
        C^{11}_* & C^{12}_* \\
        C^{21}_* & C^{22}_*
    \end{pmatrix}
    \end{pmatrix}
\end{equation}
Similarly, for the matrix $\mathbb{M}$, we have
\begin{align}
&(\mathrm{S}\mathbb{R})\mathbb{M}(\mathbb{R}^T\mathrm{S^T}) = \\
&=\begin{pmatrix}
        \begin{pmatrix}
        0 & 0 \\
        0 & 0
    \end{pmatrix} & \begin{pmatrix}
        0 & 0 \\
        0 & 0
    \end{pmatrix} \\
        \begin{pmatrix}
        0 & 0 \\
        0 & 0
    \end{pmatrix} & \begin{pmatrix}
        1 & 0 \\
        0 & 1
    \end{pmatrix}
    \end{pmatrix}\times\begin{pmatrix}
        \begin{pmatrix}
        \eta^{-1}1 & 0 \\
        0 & \eta^{-1}1
    \end{pmatrix} & \begin{pmatrix}
        0 & 0 \\
        0 & 0
    \end{pmatrix} \\
        \begin{pmatrix}
        0 & 0 \\
        0 & 0
    \end{pmatrix} & \begin{pmatrix}
        C^{11}_* & C^{12}_* \\
        C^{21}_* & C^{22}_*
    \end{pmatrix}^{-1}
    \end{pmatrix}\times\begin{pmatrix}
        \begin{pmatrix}
        0 & 0 \\
        0 & 0
    \end{pmatrix} & \begin{pmatrix}
        0 & 0 \\
        0 & 0
    \end{pmatrix} \\
        \begin{pmatrix}
        0 & 0 \\
        0 & 0
    \end{pmatrix} & \begin{pmatrix}
        1 & 0 \\
        0 & 1
    \end{pmatrix}
    \end{pmatrix} \\
    &=\begin{pmatrix}
        \begin{pmatrix}
        0 & 0 \\
        0 & 0
    \end{pmatrix} & \begin{pmatrix}
        0 & 0 \\
        0 & 0
    \end{pmatrix} \\
        \begin{pmatrix}
        0 & 0 \\
        0 & 0
    \end{pmatrix} & \begin{pmatrix}
        C^{11}_* & C^{12}_* \\
        C^{21}_* & C^{22}_*
    \end{pmatrix}^{-1}
    \end{pmatrix}
\end{align}
This completes the justification of the definition of $\mathbb{M}$. Similarly, one shows that the losslessness property holds as well.
\end{document}